\documentclass[sigplan,screen,authorversion,nonacm]{acmart}

\usepackage{amsmath}
\usepackage{amsthm}

\usepackage{url}
\usepackage{listings, lstautogobble}
\usepackage{xspace}

\usepackage{hyperref}

\newcommand{\bosque}{\textsc{Bosque}\xspace}
\newcommand{\srclocation}{\url{https://github.com/BosqueLanguage/BosqueCore}}

\newcommand{\eg}{\hbox{\emph{e.g.}}\xspace}

\newcommand{\etc}{\hbox{\emph{etc.}}\xspace}

\usepackage{color}
\definecolor{purple}{RGB}{75, 0, 255}
\definecolor{cgreen}{rgb}{0.25,0.5,0.35} % comments

\lstdefinelanguage{bosque}{
keywords={concept, entity, datatype, typedecl, enum, type, provides, field, env, switch, match, abstract, method, if, then, elif, else, function, return, true, false, none, let, var, in, requires, ensures, invariant, validate, recursive, sensitive, using, of, this, pred, fn, ref, examples, for, defer, test, const, override},
keywordstyle=\color{blue}\bfseries,
identifierstyle=\color{black},
alsoother={@},
sensitive=true,
comment=[l]{\%\%},
commentstyle=\color{cgreen}\bfseries
}
 
\lstdefinelanguage{bosquelit}{
keywords={let, in, true, false, none},
keywordstyle=\color{blue}\bfseries,
identifierstyle=\color{black},
alsoother={@},
sensitive=true,
comment=[l]{\%\%},
commentstyle=\color{cgreen}\bfseries
}
 
\setcopyright{acmlicensed}
\copyrightyear{2024}
\acmYear{2024}
\acmDOI{XXXXXXX.XXXXXXX}

\begin{document}

%%
%% The "title" command has an optional parameter,
%% allowing the author to define a "short title" to be used in page headers.
\title{An Effectively $\Omega(c)$ Language and Runtime}

\author{Mark Marron}
\email{marron@cs.uky.edu}
%\orcid{1234-5678-9012}
\affiliation{%
  \institution{University of Kentucky}
  \city{Lexington}
  \state{Kentucky}
  \country{USA}
}

%%
%% The abstract is a short summary of the work to be presented in the
%% article.
\begin{abstract}
The performance of an application/runtime is usually thought of as a continuous function where, the lower the amount of memory/time used on a 
given workload, then the better the compiler/runtime is. However, in practice, good performance of an application is conceptually 
more of a binary function -- either the application responds in under, say $100$ms, and is fast enough for 
a user to barely notice~\cite{usability}, or it takes a noticeable amount of time, leaving the user waiting and potentially abandoning the task. Thus, performance 
really means how often the application is fast enough to be usable, leading industrial developers to focus on the $95$th and $99$th percentile 
latencies as heavily, or moreso, than average response time.

Unfortunately, tracking and optimizing for these high percentile latencies is difficult and often requires a deep understanding of the application, 
runtime, GC, and OS interactions. This is further complicated by the fact that tail performance is often only seen occasionally, and is specific to 
a certain workload or input, making these issues uniquely painful to handle. Our vision is to create a language and runtime that is designed to be 
$\Omega(c)$ in its performance -- that is, it is designed to have an \underline{effectively constant time} to execute all operations, there is a \underline{constant 
fixed memory} overhead for the application footprint, and the garbage-collector performs a \underline{constant amount of work} per allocation + a (small) 
\underline{bounded pause} for all collection/release operations. 

\end{abstract}

%%
%% The code below is generated by the tool at http://dl.acm.org/ccs.cfm.
%% Please copy and paste the code instead of the example below.
%%
%\begin{CCSXML}
%<ccs2012>
% <concept>
%  <concept_id>00000000.0000000.0000000</concept_id>
%  <concept_desc>Do Not Use This Code, Generate the Correct Terms for Your Paper</concept_desc>
%  <concept_significance>500</concept_significance>
% </concept>
%</ccs2012>
%\end{CCSXML}

%\ccsdesc[500]{Do Not Use This Code~Generate the Correct Terms for Your Paper}

%\keywords{Keywords!!!}

%%
%% This command processes the author and affiliation and title
%% information and builds the first part of the formatted document.
\maketitle

\section{Introduction}
A key-performance-indicator (KPI) for many applications is the $99$th (or $95$th) percentile latency -- that is, the time it takes for the application to respond 
to a user request $99\%$ of the time. This is a critical metric as these \emph{tail-latency} events are often pain points for users and, once encountered, lead 
to disengagement~\cite{usability}. Unfortunately, these tail-latency events are often intermittent, involve multiple events, and even different subsystems~\cite{tailatscale,toddmeasure}. 
These features combine to make them very difficult to diagnose and resolve.

Some sources of tail-latency are irreducible parts of a distributed (or networked) application, such as connection latency, shared resource contention, or intermittent failures. However, 
the latency from these sources is often amplified by runtime and application behavior. For example, a network stall that leads to requests backing up, which leads 
to many objects being promoted into old GC generations, leading to a long GC pause during whole heap collection, causing more requests to back up, and so on. As seen in this 
example, the triggering event for the latency spike is an intermittent network stall but the amplification, along with triage work and resolution, is in the runtime behavior. 

Our vision, and work-in-progress, is to create a language and runtime that is designed to be $\Omega(c)$ in its performance and memory use behaviors -- that is, it is designed to have an
effectively constant time to execute all operations, there is a constant fixed memory overhead for the application footprint, and the garbage-collector performs a
constant amount of work per allocation + a (small) bounded pause for all collection/release operations. This is a radical departure from the current state of the art 
where modern runtimes struggle with work-case behaviors and heuristics, standard libraries often have pathological corner cases~\cite{redos,v8dos}, and even state-of
the art GCs have degenerate slow-paths for certain scenarios~\cite{distillingcost}.

This work-in-progress paper looks at a novel programming language, \bosque~\cite{bosque,highassurance,bsqon}, which has a number of critical features that we believe will allow us to achieve 
this vision. In particular the language and standard libraries are designed with the following core principles:
\begin{itemize}
  \item \textbf{Fully Defined and Deterministic:} The language does not have any compiler or runtime \emph{implementation definable} or \emph{non-deterministic} behavior -- that is every program always has a fixed and unique result.
  \item \textbf{Immutability:} All values in the language are immutable and cannot be changed after creation.
  \item \textbf{No Cycles:} There are no cycles in the object graph and no way to create them.
  \item \textbf{No Identity:} There is no address or pointer comparison and no way to observe the identity of an object, \eg the language is fully referentially transparent.
  \item \textbf{Asymptotically Stable Algorithms:} The language and core libraries use only algorithms that have a constant or $\Omega$ bounded linear/logarithmic cost instead of amortized constant 
  (but worst case quadratic) costs.
\end{itemize} 

Using these features, this paper presents a discussion of the core runtime and compilation design choices that we believe will allow us to achieve our vision for an $\Omega(c)$ runtime 
and how it will address the internal components of tail-latency issues in a foundational way.

%%\section{\bosque Language Discussion}

\section{Core Runtime and Compilation}
The design and implementation of the standard library and runtime contribute to many (smaller) performance variance issues and, in some cases, are the source of 
major pathological behaviors. In this section we cover some of the widely occurring scenarios and how \bosque makes alternative design choices that, at the cost of some maximum performance 
in the best-case, lead to much better worst-case performance and a more stable performance overall.

\subsection{Standard Library Data Structures}
Standard data-structures are a common location for high-variance algorithms to appear. In particular, many classic and average-case efficient data structures, are 
best (and amortized) case $O(1)$ but have slow-paths that are $O(n)$. For example, the classic open-addressed hashtable is $O(1)$ for lookup and insertion but can be $O(n)$ 
when re-hashing is needed. This popular implementation (Java, C\#, V8) is a classic example of a high-variance data-structure that may 
perform well under normal conditions (and load testing) but contributes to tail-latency issues or even Denial-of-Service (DoS) attacks~\cite{v8dos} in production systems. 

Other common data-structures include sets, often hash-table backed, strings with amortized re-layout, and vectors with amortized resizing. In all of these cases 
there is an amortized algorithm with a fast path (often $O(1)$) and an amortized slow path (often $O(n)$) that is triggered by some condition. In all of these cases 
there are alternative data-structures that have higher average but lower worst-case performance, usually $\Omega(log~n)$ or effectively $O(1)$ in practice~\cite{persistentrrb}. 
\bosque has chosen to use these alternative data-structures in the standard library to trade some amount of best-case performance for a lower variance in general 
performance.

\subsection{Standard Library Algorithms}
Another source of variance and bugs in modern runtimes is the use of high-variance and non-deterministic algorithms in the standard library. Sorting with 
quick-sort, is unstable and $O(n~log~n)$ with $O(n^2)$ worst-case, and a painful source of bugs~\cite{jssort} is a classic example. As with data-structures, we can 
replace these algorithms with slightly less performant best case versions, \eg quicksort with timsort as V8 has done or moving from unspecified enumeration 
orders to fixed in maps/sets, to get more stable performance overall -- and fewer surprised developers as a bonus! 

\subsection{Regular Expressions}
As a source of major DoS vulnerabilities and worst-case exponential behaviors from benign looking code, regular expressions are a key source of standard library 
functionality that can cause major issues in production systems. The classic example is pathological backtracking that can occur with certain patterns and, if 
reachable from user input, can lead to crippling DoS attacks~\cite{redos}. This is a complex issue as, no known algorithm exists for handling 
PCRE regular expressions that is immune to this issue. Instead we have chosen to rework the regular expression language in \bosque~\footnote{The BREX regex 
language and runtime are available at: \url{https://github.com/BosqueLanguage/BREX}} to ensure that the semantics match a polynomial NFA simulation implementation. 

The design of the \bosque regular expression language, BREX, starts with a classic definition of a regular language and adds support for limited forms of lookahead/behind 
along with stratified negation and conjunction. This design allows the expression of most regex uses in practice~\cite{regexuse}, while ensuring that worst case behavior is at most 
polynomial and avoids the pathological ReDoS behavior in backtracking based engines. Large polynomial execution times are still possible, these are much less likely to be an issue 
in practice~\cite{redos}.

\subsection{Compilation and Caching}
Language semantic choices can have a non-trivial impact on runtime/compiler implementation options and possible sources of variance. For example, JavaScript is a highly 
dynamic language that relies on aggressive JIT use and caching to achieve good performance. This is a source of variance in the application as performance can vary 
widely as JIT code warms up and/or changes modes, changes to JIT heuristics can result in large performance swings for certain workloads~\footnote{Particularly in the 
presence of the \emph{JIT specific} coding and heuristic assumptions.}, and bad luck around behavior, \eg polymorphic inline caches~\cite{polycache}, can cause hard to diagnose performance issues.

The \bosque language is designed to support static compilation and resolution whenever possible. The semantics of the language defaults to closed inheritance except where 
explicitly marked, there are no ur-classes or prototype chains with dynamic dispatch, and the type system does not allow the creation or ad-hoc extension of dynamic types/calls. 
Most critically, the compilation model of an assembly is fully closed -- so aggressive compile time tree-shaking and name resolution are possible. This allows the 
direct and complete elimination of many sources of dynamic behavior related variance.

However, \bosque still supports dynamic dispatch and multiple inheritance so full elimination or execution lookups, or even reduction to always known offsets, is not possible. 
In these cases we believe that \bosque still provides a unique opportunity as, at each call site, the full set of possible targets is known and can be used to generate a 
fixed strategy for lookup. The compiler can gradually apply more expensive lookup strategies as the number of targets grows, starting with simple inline cases, moving to (SIMD) 
accelerated linear lookups, and finally to fully general resolution for the largest number of targets. This allows the runtime to have a fixed and predictable cost for 
each call site and, if possible, can hoist redundant resolution at compile time.

\section{GC Algorithm and Defragmentation}
The garbage collector is a critical component of a runtime system and is often a major source of variance in application performance behavior. Massive work has gone into various 
GC algorithms to reduce their, costs, pause times, and memory overheads~\cite{gchandbook}. A particular focus in recent years has been on reducing pause times~\cite{chickenclover,lxr}. 
However, in a language with mutation, cycles, and semantically observable object identity, there are fundamental limitations to what can be achieved -- specifcally tradeoffs 
between latency and throughput~\cite{distillingcost} and the increasing complexity of the memory management implementation~\cite{lxr}.

\bosque presents a unique opportunity to re-think GC design and implementation. The language is designed with fully immutable values/objects, no cycles, and no way to 
observe identity (directly via addresses or indirectly via any language semantics). This allows for novel and aggressive design choices to be made in the GC and allocator 
implementation with the aim of achieving the following:
\begin{itemize}
\item \textbf{Constant Memory Overhead:} The memory consumed should be bounded by $K + M \times \emph{\#objs}$ where \emph{\#objs} is the number of 
allocated objects, $K$ is a fixed overhead, and $M$ is a small per-allocation overhead.
\item \textbf{Fixed Work Per Allocation:} The work done by the GC for each allocation should be (effectively) constant -- regardless of the lifetime or application behavior.
\item \textbf{Bounded Collector Pauses:} The collector should only require the application to pause for a (small) bounded period.
\item \textbf{Application Code Independence:} The application code should not pay any cost, \eg write barriers, remembered sets, \etc, for the GC implementation.
\item \textbf{Defragmentable:} The GC should be able to compact the heap to manage fragmentation -- eliminating performance (and memory use) issues in long running applications.
\end{itemize}

Based on the design of a generational GC with a copying young-space and a reference-counting old-space~\cite{lxr,urc} we believe that these goals are achievable. By construction, 
and the fact that \bosque prevents cycles, this algorithm satisfies our first goal of a fixed and constant memory overhead. The size of the young generation defines the $K$ and 
the per-object metadata gives the $M$ in the formula above. As, described below, since the old generation is also defragmentable we have a fixed overhead for external fragmentation as well.

As \bosque values are immutable then once an object is copied to the old generation it will never have any pointer updates. This eliminates the need for any write barriers as well 
as any work to re-process reference counts on changes. This means that the work for any object is given by $\emph{Alloc} + R_{\text{survive}}\times(\emph{Copy} + \emph{Inc} + \emph{Dec}) + \emph{Release}$, 
all\footnote{The cost of de-fragmentation is not included here but is an asymptotically constant value.} of which are (small) constant time operations!

In addition to having a low total-cost per allocation, the design of the \bosque language also allows for us to ensure that the collector pauses are bounded too -- that is we 
are not trading off throughput for large and/or unpredictable collector latency or risking the GC failing behind and stalling the mutator~\cite{distillingcost}. In particular, 
the direct implementation is to perform a stop-the-world collection (STW) of the young generation and a constant amount of ref-count work in the old generation. This bounds 
the pause to the time to process the young generation + a constant time for the old generation while always reclaiming memory as fast as the application thread uses/recycles 
it\footnote{It may be possible to further reduce this pause with more sophisticated designs but we have not explored this yet.}.

Our prototype collector design, using the immutability and cycle-freedom of \bosque, does not require any write/read barriers or remembered sets. It even works nicely with 
conservative collection~\cite{conservativegc}, enabling the compiler to skip root-maps, and easily support pointers into the stack and interior value pointers! This is a major 
simplification of the GC design and, of course, allows for more aggressive runtime design and compiler optimizations.

Finally, the defragmentation of the old generation is a critical feature for long running applications. Traditionally, moving objects in the old generation is a major challenge as 
it requires updating all pointers to the object in an atomic fashion~\cite{chickenclover} to maintain the semantics of the program. This is, of course, a complex and potentially 
expensive operation. However, the \bosque language design ensures that all values are truly referentially transparent and there is no way to observe the identity of an object. So, 
if the GC wants to move an object it can simply make a duplicate copy and, incrementally, update any references from the original version to the copy. Since, from the application semantic 
perspective, these values are indistinguishable it does not matter if the application uses the old and new version of the object simultaneously! This allows the GC to defragment the 
old generation incrementally\footnote{We can also optimize for the common case of $1$ parent by using the ref-count info to store the parent pointer and then defragmenting eagerly.} 
and without any support from the application code. 

\section{First Class Telemetry and Analytics}
Even with a $\Omega(c)$ runtime there will be $99$th percentile outliers and other performance issues, coming from operating system behavior, shared resource contention, network latency, 
or even intermittent failures, that need to be diagnosed and resolved. Traditionally, this has been addressed with a combination of $3^{\text{rd}}$ party logging, monitoring, and profiling tools 
that are manually integrated into the application code. Even though this information is eventually needed by every application and the metrics required are often similar, the integration 
is a manual, ad-hoc, and sometimes ill-integrated process. 

\bosque provides a unique opportunity to address this issue by making telemetry and analytics a first-class concern of the language and runtime. The language design allows for the 
direct integration of telemetry and analytics into the system~\cite{logpp}. Further, as \bosque constrains when and where non-deterministic interactions can occur, the runtime can provide 
automated hooks to collect a baseline of performance information and metadata for analysis. Interestingly, as \bosque is deterministically replayable, it may also be possible to 
capture traffic for replay under different conditions to explore where and what subsystems a tail performance issue may be related to.

\section{Related Work}
Prior work on the problem of tail-latency has primarily focused on issues at the network, cloud, and OS level~\cite{tailatscale,microbench,ebpf,catapult}. However, in many ways these 
issues are easier for developers to address, usually by increasing redundancy, increasing capacity, and leveraging the optimized networking/load management that cloud providers offer. The types 
of issues, arising from within a process, are much harder to diagnose and resolve~\cite{privatecommtail}. The work on this subtopic is much more limited outside of the GC 
space~\cite{distillingcost,lxr,shenandoah}, and even here the results are theoretically limited by what the language semantics allow. In contrast this work-in-progress presents a clean 
break with these traditional theoretical limitations and introduces a holistic vision for the runtime/language that can address this problem in a foundational way.

\section{Onward!}
This (work-in-progress) paper presents an outline of our thinking on one of the most pressing issues around application performance behavior, tail-latency, along with a vision of how this 
can be addressed with novel programming-language and runtime/compiler design. Based on our prior work on the \bosque language, we believe that this vision is achievable and, based on 
prototypes of the concepts in this paper, that this vision is practical. We are excited to continue this work and to explore the full implications of this design. In the end we hope that 
this work will lead to a new generation of applications that are more stable, more predictable, and more performant than ever before -- saving developers from late-nights of on-call work and 
the pain of tail-latency performance anomalies!

\section*{Data-Availability Statement}
The code for the \bosque language is open-source (MIT licensed) and available at \srclocation.

%\begin{acks}
%To Robert, for the bagels and explaining CMYK and color spaces.
%\end{acks}

\bibliographystyle{ACM-Reference-Format}
\bibliography{bibfile}

\end{document}